\documentclass[prl,twocolumn,showpacs,groupedaddress]{revtex4-1}
\usepackage{epsfig}
\usepackage{subfigure}
\usepackage{graphicx}
\usepackage{amsfonts}
\usepackage[figuresright]{rotating}
\usepackage{amssymb}
\usepackage{amsmath}
\usepackage{psfrag}
\usepackage{esint} 
\usepackage{bm}
\usepackage[colorlinks,linkcolor=blue,anchorcolor=blue,citecolor=blue,urlcolor=blue]{hyperref}

\def\be{\begin{equation}} \def\ee{\end{equation}}
\def\bea{\begin{eqnarray}} \def\eea{\end{eqnarray}}

\def\nn{\nonumber}

\def\k{{\bf k}}

\def\d{{\bf d}}
\def\bsigma{{\mathbf{\sigma}}}

\makeatletter
\newcommand*{\balancecolsandclearpage}{%
  \close@column@grid
  \clearpage
}
\makeatother

\begin{document}
\title{Berry Phase Enforced Spinor Pairing}
\author{Yi Li}
\affiliation{Department of Physics and Astronomy, Johns Hopkins University, Baltimore, Maryland 21218, USA}
\date{January 16, 2020}

\begin{abstract}
Pairing symmetry plays a central role in the study of superconductivity. 
It is usually characterized by integer partial-waves, for example, $s$-, $p$-, $d$-waves. 
In this article, we investigate a new class of topological superconductivity whose gap functions possess a half-odd-integer monopole charge and, therefore, fractionalized half-odd-integer partial-wave symmetry in three dimensions.
This exotic pairing occurs between Fermi surfaces of which Chern numbers are differed by an odd integer. 
The corresponding superconducting gap function is represented by monopole harmonics with half-odd-integer monopole charges, and thus carries spinor partial-wave symmetries.
The spinor gap function can exhibit an odd number of nodes on a closed Fermi surface, which distinguishes it from all the previously known superconducting pairing symmetry. In the presence of spatial inhomogeneity of order parameters, its superfluid velocity exhibits a fractionalized Mermin-Ho relation \cite{Mermin1976}.
\end{abstract}
\maketitle

{\it Introduction.} --
The pairing symmetry is a fundamental issue to determine the properties of superconducting states.
The discovery of new pairing symmetry is always accompanied by the establishment of a new paradigm in superconductivity. 
It has long been assumed that all superconductors are completely
classified by spherical harmonic symmetries and their lattice counterparts.
Celebrated examples of unconventional superconductors and superfluid
include $p$-wave superfluid $^3$He \cite{Anderson1961,Balian1963,Leggett1975,Volovik2003}, 
heavy-fermion superconductors \cite{Stewart1984,Sigrist1991,Pfleiderer2009},
$d$-wave high-$T_c$ cuprates \cite{VanHarlingen1995,Tsuei2000},
and $s_{\pm}$-wave iron pnictides \cite{Stewart2011,Dai2015}.
Distinct symmetry of gap functions gives rise to characteristic properties in different superconducting states. 

On another hand, exciting progress has been made in the discovery of new classes of quantum materials exhibiting topological electron bands and topological gaps which include topological insulators \cite{Kane2005a,Bernevig2006a,Bernevig2006,Fu2007a,Schnyder2008,Roy2009,Zhang2009,Xia2009,Chen2009,Hasan2010,Moore2010,Qi2011} and quantum anomalous Hall insulators \cite{Haldane1988,Yu2010,Chang2013,Haldane2014,Liu2016}. 
Further development includes the prediction and discovery of topological semi-metals \cite{Murakami2007,Wan2011,Xu2011,Yang2011,Burkov2011,Wang2012,Witczak-Krempa2012,Hosur2012,Young2012,Xu2015,Lv2015,Lu2015a,Xiong2015a,Bansil2016,Bradlyn2017,Armitage2018} which opens up the discovery of a plethora of topological materials. In both cases, topological properties are encoded in the non-trivial geometric phase of single-particle Bloch wave functions \cite{Berry1984,Xiao2010,King-Smith1993,Nagaosa2010,Kane2005a,Roy2009}, defined either over a two-dimensional (2D) Brillouin zone in an anomalous Hall insulator, or over a 2D Fermi surface in a 3D Weyl semi-metal, which are characterized by a non-zero integer-valued Chern number.

Recently, monopole harmonic superconductivity \cite{Li2018} has been proposed based on the generalization of the single-particle Berry phase to ``\textit{pair} Berry phase" -- the two-particle Berry phase for Cooper pairs in superconductors \cite{Murakami2003a}. 
This novel topological class of 3D superconductors can possibly exist in, for example, magnetic Weyl semi-metals under the proximity effect with an ordinary $s$-wave superconductor. Generally, when pairing occurs between two Fermi surfaces of opposite Chern numbers, {\it i.e.}, the enclosed Weyl points have opposite chiralities, the inter-Fermi-surface Cooper pair can inherit band topology non-trivially and acquire a non-trivial ``\textit{pair} Berry phase" in the weak-coupling regime. 
The ``\textit{pair} Berry phase", as a type of topological obstruction, prevents the gap function to be well-defined over the entire Fermi surface enclosing a Weyl point. Hence, the corresponding gap function is no longer describable by spherical harmonic functions or their lattice counterparts. Instead, it is characterized by monopole harmonic functions which are the eigenfunctions of angular momentum in the presence of a magnetic monopole \cite{Dirac1931,Tamm1931,Wu1976,Haldane1983,Li2018}. 
The ``\textit{pair} Berry phase" further enforces gap nodes and determines the total vorticity of gap nodes over a Fermi surface which is independent of specific paring mechanisms.
Therefore,  monopole harmonic superconductivity is fundamentally different from the previously known unconventional superconductivity, for example, $d$-wave high-T$_c$ superconductivity and $p$-wave $^3$He-$A$ superfluid. Furthermore, monopole harmonic superconductivity is only an example of topological many-particle order. In the particle-hole channel, monopole harmonic charge-density-wave (CDW) order has been proposed in a model of Wely semi-metal consisting nesting Fermi surfaces enclosing Weyl points of the same chirality \cite{Bobrow2018}.
This novel topological class of CDW states is also characterized by monopole harmonic symmetry and can host topologically protected emergent Weyl nodes in the gap functions. 

In this article, we first explain the difference between the familiar examples of topological superfluid/superconductors  \cite{Balian1963,Read2000,Volovik2003,Fu2008,Schnyder2008,Chung2009,Lutchyn2010,Qi2011,Alicea2012,Sato2017} and the monopole harmonic superconductor. 
Then, we investigate the spinor pairing as an exotic example of monopole harmonic pairing states. When electrons pair between two topological Fermi surfaces carrying Chern numbers with different even- and oddness, the Cooper pair acquires a half-odd-integer monopole charge. The gap function can exhibit an odd number of nodes over a single Fermi surface and leads to non-trivial Bogoliubov excitation spectrum.
The superfluid velocity obeys a fractional generalization of the Mermin-Ho relation \cite{Mermin1976} in the presence of spatial inhomogeneity of order parameters. 

{\it Monopole harmonic pairing and complex $d$-vectors.} -- We begin with examples of topological superconductivity in which the Fermi surface is topologically trivial but the superconducting gap function exhibits non-trivial topology.
For example, the spin-polarized $p_x+ip_y$ pairing is fully gapped in 2D. It belongs to the $D$-class which breaks both spin-rotation and time-reversal symmetry.
The gap function $\Delta(\mathbf{k})$ is a complex function exhibiting a phase winding number $\nu = \frac{1}{2\pi}\oint d \k \partial_{\k} \theta (\mathbf{k})$ around the 1D Fermi circle, where $\theta(\k)$ is the U(1) phase of the gap function.
For the 3D time-reversal invariant pairing of the $^3$He-B type, it belongs to the DIII class, and the single-particle band structure exhibits two-fold spin degeneracy.
The corresponding gap function is no longer a scalar function but represented by a pairing matrix.
It is proportional to a $2\times 2$ unitary matrix, say, $\Delta(\k)=i\sigma_y \hat \d(\k)\cdot \bsigma$, in which the real $d$-vector exhibits a nontrivial texture over the Fermi surface characterized by a non-trivial Pontryagin index,
\bea
\nu=\frac{1}{8\pi}\oiint_{\mathrm{FS}} d k^2 ~ \epsilon_{\mu\nu} \hat \d(\k) \cdot
\partial_{k_\mu} \hat \d(\k) \times \partial_{k_\nu} \hat \d(\k).
\eea
Under an open boundary condition, C-class 2D topological superconductors exhibit 1D chiral Majorana modes on the edge; while, DIII-class 3D time-reversal invariant topological insulators exhibit helical 2D Majorana surface modes.

However, when the single-particle band structure exhibits a non-trivial Berry
phase, the above classification scheme does not always hold for the superconductivity in this system.
A novel topological class of
superconductivity -- the monopole harmonic superconductivity has been proposed \cite{Li2018}.
The key point is that when Cooper pairing occurs between two Fermi surfaces
carrying different Chern numbers, the Cooper pairs develop a non-trivial two-particle pair Berry phase.
As a concrete example, consider the simplest model of Weyl semi-metal state with only a
pair of Weyl points located at $\pm\mathbf{K}_0$.
Upon doping, a pair of separated Fermi surfaces appear around the
Weyl points, denoted by FS$_\pm$ respectively, around which the low
energy Hamiltonians become
\bea
H_\pm (\mathbf{k} \mp \mathbf{K}_0)=\pm v_F \mathbf{k} \cdot \bsigma -\mu.
\eea
FS$_\pm$ exhibit non-trivial single-particle Berry phases and the
associated single-particle monopole charges are $\pm q$ with $q=\frac{1}{2}$,
respectively, or, the Chern numbers are $\pm C$ with $C=2q$.
The gap function of the pairing between FS$_{\pm}$ can be expressed in the 2-component
fermion basis 
which is generally represented by a $2 \times 2$ pairing matrix.
Different from the case of $^3$He-B, the monopole superconductivity is an example of \textit{non-unitary pairing}, exhibiting complex $\d$-vectors because of broken  time-reversal symmetry, 
$\frac{1}{\sqrt 2} (-d_x+i d_y)=u_k^2, ~~
\frac{1}{\sqrt 2} (d_x+i d_y)=v_k^2, d_z=\sqrt 2 u_k v_k,
$
where $u_k=\cos\frac{\theta_k}{2}$
and $v_k=\sin\frac{\theta_k}{2} e^{i\phi_k}$.
The pair Berry phase can be defined in terms of complex $\d$-vectors as
\bea
\mathbf{A}_p (\mathbf{k}) =\hat \d^* i {\mathbf\nabla_{k}} \hat \d
=2q_p \tan\frac{\theta}{2}\hat e_\phi.
\eea
Correspondingly, the associated Berry curvature can be expressed as
$
\Omega_i(\mathbf{k})=\epsilon_{ijk}\partial_j A_{p,k}
=
i\epsilon_{ijk}\partial_j \hat \d^* \cdot
\partial_k \hat \d.
$
The total pair Berry flux through S$_+$ is
\bea
\oiint_{S_+} d\mathbf{k} \cdot \mathbf{\nabla}_{k} \times
\mathbf{A}_p(\mathbf{k})=
\oiint_{S_+} d\mathbf{k} \cdot 
i\partial_j \hat \d^* \times
\partial_k \hat \d
= 4\pi q_p. \nn \\
\eea
where the pair monopole charge $q_p=2q$.
Hence, the inter-Fermi-surface pairing inherits
the Berry fluxes of single electrons from different topological Fermi surfaces in a non-trivial way.

Consequently, the topological obstruction in the wavefunction of Cooper pairs prevents the phase of its gap function to be well-defined
over the entire Fermi surface, which leads to generic nodal structures of pairing gap functions. 
The gap function $\Delta(\mathbf{k})$ projected on FS$_\pm$
possesses a generic nodal structure with a total vorticity $2q_p$,
which is independent of specific pairing mechanisms and symmetry \cite{Li2018}.
When $q_p\neq 0$, $\Delta(\mathbf{k})$ cannot be a
regular function over the entire FS$_+$.
The nodal structure of $\Delta(\mathbf{k})$ at $q_p\neq 0$ is distinct
from that of the usual pairing symmetry based on spherical harmonics
$Y_{lm}(\hat{\mathbf{k}})$, which are regular functions over a spherical Fermi surface.
The latter corresponds to $q_p=0$ and the total vorticity equals zero.
For example, for the $^3$He-A type $p_x+ip_y$ pairing, two gap nodes
lie at the north and  south poles as a pair of vortex and anti-vortex,
respectively.

Away from the Fermi surface, the nodes of
$\Delta(\mathbf{k})$ at the Fermi surface extend into vortex lines in momentum space.
The Weyl points are sources and drains of the vortex lines.
Vortex lines connecting between the Weyl points intersect a Fermi 
surface forming vortices and antivortices with a total vorticity of $\pm 2 q_p$. 
Vortex lines not connecting to the Weyl points form closed loops,
and their intersections on the Fermi surfaces form
pairs of vortices and anti-vortices.
If only $2q_p$ nodes appear, this pairing is referred to as {\it fundamental}.
If extra nodes on each Fermi surface arise 
due to the additional vortex and anti-vortex pairs,
the corresponding pairing
is referred to as {\it non-fundamental}.

{\it Emergent Weyl nodes in Bogoliubov spectrum.} -- The Bogoliubov nodal excitations are characterized by the monopole charge $q_p$.
Around each gap node, the low-energy quasi-particles Hamiltonian is
in a 3D Majorana-Weyl form as
\bea
H_{qp}=v \delta \mathbf{k} \cdot \hat n~ \tau_3 +\Delta (\delta \mathbf{k}) \tau_+
+\Delta^*(\delta \mathbf{k}) \tau_-,
\label{eq:majorana_weyl}
\eea
where $\tau_3$ and $\tau_\pm=(\tau_1\pm i\tau_2)/2$ are Pauli matrices defined
in the Nambu space;
$\hat n$ is the local normal direction
at the gap node on the Fermi surface.
The chirality index of the Hamiltonian in Eq. \eqref{eq:majorana_weyl} is determined by the nodal vorticity of $\Delta(\delta \mathbf{k})$ on the Fermi surface. 
A remarkable feature is that the low-energy nodal excitations non-trivially inherit the topology from band-structure Weyl points,
although the band Weyl points are at high energy near the cut-off scale away from the 
Fermi energy after doping. 
The philosophy of renormalization group tells us that 
low-energy physics is usually insensitive to the details at high energy.
However, due to the non-perturbative nature of topological properties,
the Weyl point determines the overall topological structure
which controls the nodal structure of $\Delta(\mathbf{k})$.

{\it Spinor pairing from half-odd-integer pair monopole charges.} -- So far, we have only discussed the monopole pairing with an integer-valued
monopole charge $q_p$, for which the pairing symmetry still lies in the
integer partial-wave channels.
Nevertheless, if we further consider the pairing between two Fermi
surfaces carrying Chern numbers with different even- and oddness,
the pair monopole charge, $q_p=\frac{1}{2} |C_1-C_2|$, is a half-odd-integer.
A remarkable property is that the above pairing order parameter,
which is bosonic, forms a spinor representation under rotation.

Consider the following system consisting of two different
types of fermions.
The first type is a spin-$\frac{1}{2}$ fermion, with its annihilation operator denoted as $c_{\alpha}$, $\alpha=\uparrow,\downarrow$ and mass $m$. It exhibits a 3D Weyl-type spin-orbit coupling as shown in $H_c^0$.
The second type of fermion, described by $H_d^0$, is a single-component fermion with its annihilation
operator denoted as $d$. It has a simple parabolic dispersion and mass $M$.
\bea
H_c^0&=&\sum_{\mathbf{k}}\sum_{\alpha,\beta=\uparrow,\downarrow} c^\dagger_\alpha(\mathbf{k})
\Big(\frac{\hbar^2k^2}{2m}-\lambda \mathbf{k}
\cdot \mathbf{\sigma}_{\alpha \beta}-\mu_1 \Big) c_\beta(\mathbf{k}), \nonumber \\
H_d^0&=&\sum_{\mathbf{k}} ~ d^\dagger(\mathbf{k})
\Big(\frac{\hbar^2 k^2}{2M} -\mu_2\Big) d(\mathbf{k}).
\eea
The synthetic spin-orbit coupling of the $c$-fermion has been
proposed to be realized in ultra-cold atom systems by employing
light-atom interaction \cite{Li2012,Anderson2012}.
The system breaks the inversion symmetry, and
exhibits the split Fermi surfaces denoted as FS$_\pm$
carrying the opposite monopole charges $q=\pm\frac{1}{2}$.
whose Fermi wavevectors $k_{f\pm}$ satisfy
$k^2_{f\pm}/2m \pm \lambda k=\mu$.

To enable the pairing between $c$ and $d$ Fermi surfaces, let's consider the simplest case that the Fermi surface of $d$-fermion, FS$_d$, matches one of the helical Fermi surfaces of the $c$-fermion, say, $FS_{c,+}$. This can be achieved by tuning the chemical potential $\mu_2$ of $d$-fermions such that $k_{f;d}=k_{f;c,+}$ is satisfied. 
Then, the Cooper pairing can occur between FS$_+$ and FS$_d$, which carry
Chern numbers 1 and 0, respectively.
The mean-field inter-Fermi-surface pairing Hamiltonian becomes
\bea
H_\Delta=\sum_{\mathbf{k}} \Delta_\alpha(\mathbf{k})
c^\dagger_\alpha(\mathbf{k}) d^\dagger(-\mathbf{k})
+\Delta_\alpha^*(\mathbf{k})
d (-\mathbf{k})c_\alpha(\mathbf{k}). \ \ \
\eea
The gap function exhibit a two-component spinor structure $\Delta(\mathbf{k})=(\Delta_\uparrow(\mathbf{k}),
\Delta_\downarrow(\mathbf{k}))^T$.
This system maintains rotation symmetry. Hence, its gap function
can be expanded as $\Delta(\hat k)=\sum_{jj_z} \Delta^{jj_z}(\mathbf{k})$
where the partial-wave channels are denoted by
half-integer angular momentum quantum numbers $j, j_z$.
Due to the inversion symmetry breaking, each partial-wave channel
is represented as a mixture of two channels with even and odd parity,
respectively as,
$\Delta_\alpha^{jj_z}(\mathbf{k})=\Delta_\alpha^{jj_z;l}(\mathbf{k})
+\Delta_\alpha^{jj_z;l+1}(\mathbf{k})$,
where
\bea
\Delta_\alpha^{jj_z;l}(\mathbf{k}) =\Delta_{jj_z;l}
\phi_{jj_z;l\alpha}(\hat k),
\eea
where $\phi_{jj_z;l\alpha}(\hat k )=\sum_{l_z} \langle jj_z|ll_z \frac{1}{2}
\alpha\rangle Y_{ll_z} (\hat k)\otimes |\alpha\rangle $
is the spin-orbit coupled spherical harmonic function.

For later convenience, we define the helical basis $|\lambda_+(\mathbf{k})\rangle$ which satisfy $\sigma \cdot \hat k|\lambda_\pm(\mathbf{k})\rangle =\pm |\lambda_\pm(\mathbf{k})\rangle$.
After projected to the positive helicity sector
$|\lambda_+(\mathbf{k})\rangle$,
the angular momentum $\mathbf{J}=\mathbf{L} +\frac{\sigma}{2}$ becomes
\bea
\mathbf{J}^{+}=\hat{\mathbf{k}} \times (-i \mathbf{\nabla_k}-\mathbf{A}_k) +\frac{1}{2}\hat{\mathbf{k}},
\eea
where $\mathbf{A}_k=i\langle\lambda_+|\nabla_k|\lambda_+\rangle$.
Similarly, the spin-orbit coupled harmonic functions are also
projected as $P_+ \phi_{j, j_z;l\alpha}(\hat k )=
-P_+ \phi_{j, j_z;l+1 \alpha}(\hat k )=
\frac{1}{\sqrt 2} Y_{q,jj_z}(\hat k)$,
where $q=-\frac{1}{2}$ is the monopole charge, $j=l+\frac{1}{2}$;
$Y_{q,jj_z}$ is the monopole harmonic function satisfying
$(J^{+})^2 Y_{q,jj_z}=\hbar^2 j(j+1) Y_{q,jj_z}$, and
$J_z Y_{q,jj_z}= \hbar j_z Y_{q,jj_z}$.
After states being projected to the low- energy sector near FS$_+$, 
the gap function $\Delta^{jj_z;l}(\mathbf{k})$ becomes
$\Delta^+ (\hat k)=\Delta Y_{-\frac{1}{2},jj_z}(\hat k)$,
which exhibits singular behavior over the Fermi surface. 
Similarly, we define $\chi_+^\dagger(\mathbf{k})=
\sum_{\alpha=\uparrow,\downarrow}
\lambda_{+,\alpha}(\hat{\mathbf{k}}) c^\dagger_\alpha(\mathbf{k})$.
The projected pairing Hamiltonian becomes
\bea
H^P_\Delta(\mathbf{k}) =\Delta Y_{-\frac{1}{2},jj_z}(\hat k)
\chi^\dagger(\mathbf{k}) d^\dagger(-\mathbf{k}) +h.c.
\eea

We emphasize that the projected gap function exhibits the half-integer
monopole harmonic symmetry which is independent of specific 
pairing interactions.
For example, consider the simplest  contact attractive interaction
between these two types of fermions as
\bea
H_{int}=- g \int d \mathbf{r} c^\dagger_\alpha(\mathbf{r})
 d^\dagger(\mathbf{r}) d(\mathbf{r}) c_\alpha(\mathbf{r}),
\eea
which would only give rise to the conventional $s$-wave pairing
symmetry for the topological trivial Fermi surfaces.
In this case, before projection, $\Delta_{\alpha=\uparrow,\downarrow} =-\frac{g}{V}\int
d\mathbf{r} \langle G| d(\mathbf{r}) c_\alpha(\mathbf{r}) |G\rangle$
is a constant independent of $\hat k$.
Nevertheless, after the projection, the gap functions
$\Delta_{\alpha = \uparrow,\downarrow}$ become
\bea
P^+ \Delta_{\uparrow}&=&  \Delta
Y_{-\frac{1}{2},\frac{1}{2}\frac{1}{2}} (\hat k)
= \Delta \cos\frac{\theta_k}{2}, \nn \\
P^+ \Delta_{\downarrow}&=&\Delta
Y_{-\frac{1}{2},\frac{1}{2},-\frac{1}{2}} (\hat k).
=\Delta \sin\frac{\theta_k}{2} e^{i\phi_k},
\eea
which are time-reversal partner to each other.
The projected gap functions exhibit a single point node at 
the south pole and a single node at the north pole on the FS$_+$ for the cases
of $P^+\Delta_{\uparrow}$ and $P^+\Delta_{\downarrow}$, respectively.
Again we can define the gauge invariant ``velocity" in
momentum space $\mathbf {v} (\mathbf{k})= \nabla_k \phi(\mathbf{k})-\mathbf{A}_p(\mathbf{k})$.
The circulation around the gap function nodes in both cases show
that the total vorticity
\bea
\frac{1}{2\pi} \oint_{C} d\mathbf{k} \cdot \mathbf{v}=2q=-1,
\eea
which is determined by the monopole charge.

Next we study the Bogoliubov quasi-particle excitations.
Let us consider the case of $q=-\frac{1}{2}$, $j=j_z=\frac{1}{2}$
as an example.
In the Nambu basis of $\psi(\mathbf{k})=(\chi_+(\mathbf{k}),d^\dagger (-\mathbf{k}))^T$,
we have
\bea
H(\mathbf{k})= \epsilon_k \tau_3 +\Delta \cos\frac{\theta_k}{2}
\tau_1.
\eea
The Bogoliubov quasi-particle excitations are expressed as
$
\gamma^\dagger_1 (\mathbf{k}) = \cos\frac{\beta_k}{2}
\chi^\dagger(\mathbf{k})+\sin\frac{\beta_k}{2} d(-\mathbf{k})$,
$
\gamma^\dagger_2 (\mathbf{k})=\sin\frac{\beta_k}{2} d(\mathbf{k})
- \cos\frac{\beta_k}{2} \chi^\dagger(\mathbf{-k}),
$
where $\tan \beta_k =\Delta\cos\frac{\theta_k}{2}/\epsilon_k$.
The corresponding excitation spectra are
$E^2_1(k)=\epsilon_k^2+\Delta^2 \cos^2\frac{\theta_k}{2}$,
and
$E^2_2(k)=\epsilon_k^2+\Delta^2 \sin^2\frac{\theta_k}{2}$,
which exhibit excitation nodes at the south and north
poles, respectively.

Now let us focus on the simplest case of the fundamental monopole harmonic
pairing symmetry, and consider the situation with inhomogeneous spatial
distribution of the internal symmetry of the gap function.
Before the projection, the gap function is expressed as
$\Delta(\mathbf{r})=|\Delta(\mathbf{r})|e^{i\phi(\mathbf{r})}\eta(\mathbf{r})$,
where $\eta(\mathbf{r})$ is a spin-$\frac{1}{2}$ spinor.
In momentum space, $\eta$ is projected to the monopole harmonic
function $P_+\eta = \eta_\alpha Y_{-\frac{1}{2},\frac{1}{2},\alpha}(\hat k)$.
Through the Hopf map, the spinor gap function maps to a 3-vector as
$\hat n(\mathbf{r}) =\eta^\dagger \vec \sigma \eta$.
In the case of $p$-wave superfluid of $^3$He-A phase, the direction of the
Cooper pairing orbital angular momentum is denoted by the $l$-vector,
and the curl of the superfluid velocity is determined by the spatial
variation of the $l$-vector via
$(\nabla \times \nabla\phi(\mathbf{r}))_i
=\epsilon_{ijk} \hat l \cdot \partial_j \hat l \times \partial_k
\hat l$, which is the celebrated Mermin-Ho relation \cite{Mermin1976}.
Here, because of the spinor pairing order parameter, the corresponding Mermi-Ho relation is fractionalized. The spatial variation of 
$\hat n$ leads to the following non-trivial circulation of superfluid velocity
\bea
(\nabla \times \nabla\phi(\mathbf{r}))_i =
\frac{1}{2}\epsilon_{ijk} \hat n \cdot \partial_j \hat n\times \partial_k
\hat n.
\eea
In a spherical harmonic trap, there will appear a single vortex
on the boundary induced by geometric curvature.

In summary, we have studied a class of topological nodal superconducting
states characterized by non-trivial two-particle pairing Berry phase
structure. 
The low-energy gap function symmetry is described by monopole
harmonic functions, not only in the integer-monopole charge channels,
but also in the more exotic half-odd-integer monopole charge channel. 
They can exhibit half-integer spinor partial-wave symmetry, even though
the Cooper pairs themselves are bosonic. 
We also discussed the possible experimental realizations in 
ultra-cold atom systems with synthetic spin-orbit coupling,
when Cooper pairing takes places between two Fermi surfaces
carrying Chern numbers with opposite even- and oddness. 
When the spinor pairing order parameter has  spatially
gradient, the superfluid velocity ceases to be irrotational, 
and a fractional version of the Mermin-Ho relation is derived.

{\it Acknowledgments}
Y.L. acknowledges the support by the U.S. Department of Energy, Office of Basic Energy Sciences, Division of Materials Sciences and Engineering, Grant No. DE-SC0019331 and the support by the Alfred P. Sloan Research Fellowships.

\bibliographystyle{apsrev4-1}
\bibliography{all}

\end{document}